\documentstyle[aps,preprint]{revtex}
\begin{document}
\title{Steady state properties of a driven granular medium}
\author{Gongwen Peng\thanks{Address after October 5, 1997: Department of Physics and Astronomy, Bowling Green State University, Bowling Green, OH 43403}
%\thanks{e-mail: peng@phys.ocha.ac.jp. Address after October 5, 1997: Department of Physics and Astronomy, Bowling Green State University, Bowling Green, OH 43403}
and Takao Ohta }
%\thanks{e-mail: ohta@phys.ocha.ac.jp} }
\address{Department of Physics, Ochanomizu University, Tokyo 112, Japan}
%\date{today}
\maketitle
\begin{abstract}
\indent We study a two--dimensional granular system where external driving force is applied to each particle in the system in such a way  that the system is driven into a steady state by balancing the energy input and the dissipation due to inelastic collision between particles. The velocities of the particles in the steady state satisfy the Maxwellian distribution. We measure the density--density correlation and the velocity--velocity correlation functions in the steady state and find that they are of power--law scaling forms. The  locations of collision events are  observed to be time--correlated and such a correlation is described by another power--law form.  We also find that the dissipated energy obeys a power--law distribution. These results indicate that  the system evolves into a critical state where there are neither characteristic spatial nor temporal scales in the correlation functions.  A test particle exhibits an anomalous diffusion which is apparently similar to the  Richardson law in a three--dimensional turbulent flow. 

\vspace{1.0cm}
\end{abstract}
\section{Introduction}
Granular media have distinct behaviors from  those of usual solids, fluids and gases \cite{Jaeger,Hansen,Mehta}. There are two particularly important features that contribute to the unique properties of granular materials: ordinary thermal fluctuations play no role because of the mesoscopic size of grains, and the interactions between grains are dissipative \cite{Jaeger}. Therefore, in order to maintain the dynamics of granular media in the long run, external driving forces are inevitable. Both experiments and computer simulations show that the dynamical responses of granular media to  external forces  exhibit a wide variety of interesting phenomena \cite{Jaeger,Hansen,Mehta}. Traditionally, external energy flows into the granular system from the boundaries, either by shear \cite{Louge,Campbell,Babic} or by other means \cite{lui,Bernu,Kadanoff}.  

The behaviors of inelastic and elastic systems under the same situation are quite different. Elastic systems, if contacted with a heat bath where energy flows into the system from the boundaries, will attain a homogeneous state with a uniform temperature (average kinetic energy) field. However,  the local kinetic energy  of granular systems exhibits  spatial gradients due to inelastic collision if energy is input from the boundaries \cite{Kadanoff}.  A freely evolving granular medium may have uniform temperature, but that temperature is time--dependent \cite{Deltour,Ernst}. 

A uniform and time--independent  temperature  is essentially required by the granular thermodynamics theories \cite{Edwards,Hans,Hayakawa}  or by any effort to make a comprehensive comparison between granular materials and fluids at {\it equilibrium}. Recently there are investigations  \cite{Warr,Ristow,Ichiki} showing similarity between the steady states of granular materials  and   usual fluids at equilibrium.   

Thus in order to study the properties of granular media free from the nonuniformity of temperature and anisotropy due to gravity, one needs to introduce an idealized system. In this paper, we carry out computer simulations in two dimensions of a model system satisfing the above conditions. The driving force is applied to each particle so that the external energy flows into the system uniformly.

Our main concern is the correlations in the steady state of the granular system where energy dissipation due to inelastic collision is balanced with the energy input. The spatial distribution of collision events, particle diffusion, density--density correlation and velocity--velocity correlation will be investigated by simulations. The results show that there are no characteristic spatial and temporal scales in the correlation functions, no characteristic energy scale in dissipation, and an anomalous diffusion of a test particle. 

We admit that the way of  energy input uniformly to each particle in our model is difficult to realize in experiments. As we mentioned above, however, this kind of idealization is necessary to explore the properties inherent to systems with inelastic interations and to clarify the essential difference between granular materials and usual fluids.

This paper is organized as follows. In Section II  we present the model. Numerical 
results from the model are given in Section III and we close the paper with Section IV which is devoted to discussion. 

\section{The Model}

We consider  $N$ hard disks (particles) of diameter $\sigma$  interacting via collisions in a two--dimensional square cell of length $L$ with periodic boundary conditions in the x-- and y--directions. During a collision, the relative velocity of two particles in the normal direction is reduced by  the restitution coefficient $R$ while in the tangent direction the collision is elastic:
\begin{equation}
\label{col}
\begin{array}{rcl}
{\bf  v}_i' &=& {\bf  v}_i - \frac{1}{2} (1+R) [\hat{{\bf  n}} \cdot ({\bf  v}_i - {\bf  v}_j ) ] \hat{{\bf  n}},\\[2mm]
{\bf  v}_j' &=& {\bf  v}_j + \frac{1}{2} (1+R) [\hat{{\bf  n}} \cdot ({\bf  v}_i - {\bf  v}_j ) ]  \hat{{\bf  n}}
\end{array}
\end{equation}
where primes indicate velocities after the collision and $\hat{{\bf  n}}$ is a unit vector pointing from the center of $i$th particle  to that of $j$th particle. The loss of energy due to the collision is given by
\begin{equation}
\label{eng_loss}
\delta E^- = \frac{1}{8}m(1-R^2)[\hat{{\bf  n}} \cdot ({\bf  v}_i - {\bf  v}_j ) ]^2
\end{equation}
where $m$ is the mass of the particles.

On the other hand, energy is input to the system  after each time period of $T$ in the following way: the velocities are perturbed instantaneously by a random amount:
\begin{equation}
\label{heating}
{\bf  v}_i' = {\bf  v}_i + \mbox{\boldmath $\xi$}_i
\end{equation}
where primes refer to the velocities after energy input.  The components of \mbox{\boldmath $\xi$},  $\xi_x$ and $\xi_y$,  are  random numbers distributed uniformly between $-\delta V$ and $\delta V$. The energy input rate  is therefore proportional to $(\delta V)^2/T$. The way of adding energy to the system in Eq. (\ref{heating}) can remove any macroscopic flow of the system since there is no preferable direction chosen. 

One--dimensional version of the present model was studied by Williams and MacKintosh \cite{Williams} with continuous energy input. Here the energy is added to the system instantaneously after each time period of $T$. Therefore, we are able to use the event--driven algorithm \cite{Allen,lui} to perform the simulations. An event is defined as a change of velocity either by collision or by energy input. 

The inelastic collapse singularity was observed in simulations of  inelastic system \cite{McNamara}. This is caused by  the appearance of infinite number of collisions between a group of particles which are spontaneously arranged on a straight line \cite{McNamara}. 
We avoid this singularity by a slight modified collision rule as introduced by Deltour and Barrat \cite{Deltour}. At each collision, the relative velocity  of the two particles  is first computed (Eq. (\ref{col})) and then rotated randomly by less than 5 degrees \cite{Deltour}.   

The parameters in our systems are the restitution coefficient $R$, number of particles $N$, system size $L$ and average density  (area coverage ratio) $\rho$. The diameter $\sigma$ of particles is related to them by $\rho = \frac{\pi N \sigma^2}{4 L^2}$. We will keep $L=1$ throughout the simulations, and show the results for $\delta V = 0.001$ and $T = 0.002$ unless stated otherwise. The heating period of $T=0.002$ corresponds to adding energy to the system about every five collision events for $R=0.1$. 
 
\section{Numerical Results}

\subsection{Approach to steady state}

Starting from any initial configuration the system with dissipation ($R < 1$) reaches a steady state in the center of mass frame. Since the energy input (Eq.(\ref{heating})) does not guarantee conservation of the total momentum, we subtracted the average velocity from  the velocity of each particle  in order to remove the motion of the center of mass of the whole system. Thus the velocity is defined in the frame moving with the center of mass as it was done in Ref.~\cite{Williams}. 

Fig.~1 shows how the average energy per particle relaxes with time towards a steady state for  a system with $N=1024$, $\rho = 0.16$, and $R = 0.8$.   It is found that the relaxation is exponential as $E(t) = E(\infty) + (E(0) - E(\infty))e^{-t/\tau}$.  We find that the relaxation time $\tau$ ($=2.63$ in Fig.~1) is  independent of the initial state and is controlled by the model parameters.   

Theoretically, we can derive the energy relaxation equation as follows. The system increases its energy as a result of external driving (Eq.(\ref{heating})) with an amount of 
\begin{equation}
\Delta E^{+} = \frac{1}{2} \sum_{i=1}^{N} ({\bf v}_i + \mbox{\boldmath $\xi$}_i )^2  -    \frac{1}{2} \sum_{i=1}^{N} {\bf v}_i^2 = \frac{1}{2} \sum_{i=1}^{N}  (\mbox{\boldmath $\xi$}_i^2 + 2 {\bf v}_i \cdot \mbox{\boldmath $\xi$}_i)
\end{equation}
while it decreases its energy at a collision according to Eq.(\ref{eng_loss}). The energy change rate is 
\begin{equation}
\label{relax}
\frac{\Delta E}{\Delta t} = a \Delta E^{+} - b \Delta E^{-} 
\end{equation}
where $a$ and $b$ are constants, depending on the frequency of energy input, number of particles and collision rate.

After averaging over a time interval much longer than the collision time (which will be defined later), Eq. (\ref{relax}) becomes
\begin{equation}
\label{relax2}
\frac{d E}{d t} = a' - b' E
\end{equation}
where $a' = < \frac{1}{2} a \mbox{\boldmath $\xi$}^2 > = \frac{1}{6}a(\delta V)^2$, $b' = \frac{1}{8}m(1-R^2) b$ and we have made use of $<{\bf v}_i \cdot \mbox{\boldmath $\xi$}_i> = 0$. A steady state with $E(\infty) = a'/b'$ exists for Eq. (\ref{relax2}). The time--dependent relaxation is given by
\begin{equation}
E(t) = E(\infty) + A e^{-b' t}
\end{equation}
where $A$ is an integral constant depending on the initial condition.

\subsection{Collision rate}

In the steady state the collision rate is constant, independent of time. This is in constrast to the time dependent collision rate in the freely evolving granular medium  where the total number of collisions increases as $\ln(1+t/t_e)$ for homogeneous cooling \cite{Deltour,Ernst}. In our simulations, we count the total number ($C$) of collisions  starting from a configuration in the steady state. Fig.~2(a) shows that the total collision number is linear with time for all systems with $R < 1$. The slopes of the lines are the collision rates for different values of $R$. One would  expect that  the collision rate should be an increasing function with increasing $R$ as the velocity (therefore granular temperature defined below) is larger and the collision is more likely to happen in the larger $R$ system. However, we find from Fig.~2(a) that the collision rate increases as the degree of dissipation increases. Fig.~2(b) is the dependence of the collision rate $\gamma$ as a function of degree of dissipation $\epsilon = 1 - R^2$. We find that the numerical data can be fitted using the following form
\begin{equation}
\gamma = \gamma_0 + \frac{\gamma_1}{(1-\epsilon)^c}
\end{equation}
where the exponent $c$ is found to be about $0.63$ for the data in Fig.~2. It is evident that the collision rate will diverge if $\epsilon$ is equal to $1$. 

The increase of collision rate with decreasing $R$ may be linked with the clustering mechanism found in  Refs.~\cite{Louge,Zanetti}. However, by direct visualization the density nonuniformity is not as clear as in the situation of the freely evolving granular medium \cite{Louge,Zanetti}. Fig.~3(a) is a typical configuration taken from the simulation. To find out any correlation in the collision events, we therefore resort to the following method. We record the positions where dissipation takes place by, e.g., taking the middle point's coordinates between  each colliding pair. At each dissipation place, we draw a data point and by doing so for some time interval we obtain Fig.~3(b). We see clearly that there is clustering in the dissipation places. Note that the clustering is not due to limited data points (or short time observation). In fact, there are $20,480$ points in Fig.~3(b) with overlap. We have  confirmed this observation by longer runs. It is also verified that such a clustering phenomenon does not exist for elastic system without dissipation ($R = 1$).  We  observe that the clustering patterns  change their positions and forms as  time develops.  The clustering will be  quantitatively analysed in the next subsection.  

\subsection{Time correlation of locations of dissipation}

We now determine the time correlation of the locations of dissipation places. A dissipation (a collision) takes place  in space $\boldmath r$$(t) = (x(t), y(t))$ at time $t$.  We calculate the power spectrum  of $x(t)$ and that of $y(t)$ and find that they are of power--law decaying form. Since there is no difference between the spectra of $x(t)$ and  of $y(t)$, we make average over them to obtain $P(f)$. Fig.~4(a) shows the power spectrum for $R = 0.1$, displaying the form of $P(f) \sim f^{- \alpha}$ with $\alpha = 0.49 \pm 0.01$.   The non--zero exponent reflects that clustering occurs during the dynamical process: the places where dissipation happens are more likely to be the locations for future dissipation. In this figure, the time period  corresponding to a frequency ($f$) is $t = \frac{f_{max}}{\gamma f}$ where $\gamma$ is the collision rate in the system. The upper cutoff of frequency  is $f_{max} = 1024$ and the collision rate $\gamma = 2438$. This cutoff of  frequecy in Fig.~4(a) corresponds to the time interval between two successive collisions (its inverse is the collision rate). Below that cutoff, there is no characteristic time scale in the clustering mechanism as indicated by the power--law scaling. This spectrum was obtained from the same process as shown in Fig.~3(b). We also note that the exponent $\alpha$ depends on the restitution coefficient $R$.   As $R$ increases to $1$,  $\alpha$ decreases to zero.  Fig.~4(b) shows the dependence of the exponent $\alpha$ on the restitution coefficient. As $R$ increases, the clustering becomes weaker so that the location of dissipation places becomes spatially   more uniform. This is consistent with the real space picture as shown in Fig.~3(b). 
  
\subsection{Velocity distribution}

The velocity distribution is found to be Gaussian for all the cases. Fig.~5 shows a typical distribution of the square of velocity in a semi--logarithmic plot. The linearity indicates that the velocity satisfies the Gaussian distribution like the velocity distribution in usual fluids and gases at equilibrium. We will show in Fig.~7(c) that the energy density ($e({\bf r}) = \sum_{i=1}^{N} \frac{1}{2}m{\bf v}_i^2 \delta({\bf r} -{\bf r}_i)$) has no spatial correlation, as expected from the uniform and random energy input mechanism. This means that the Gaussian distribution of velocity is valid for any part of the system. In such a situation we can now define a granular temperature $T_g = <{\bf v}_i^2>$, a quantity that is uniform in the system.  It should be noted that if the velocity does not satisfies the Gaussian distribution the variance of the velocity does not have correspondence to the usual temperature but is just internal energy. From the statistical physics' point of view, the Gaussian velocity distribution (Maxwellian distribution) comes from a very general consideration. In a conservative system, if the momentum enters into the Hamiltonian only via the kinetic energy term, the Gaussian distribution of velocity is valid irrespective of the potential energy form \cite{Landau} and the average kinetic energy is $k_BT$ with  $k_B$ the Boltzmann constant.  

Non--Gaussian velocity distribution has been found in granular materials under shear \cite{shear} and in fluidized beds of granules \cite{ichiki}.  We also find that only by uniformly energy input as in the present model  without uniaxial force like gravity can the granular materials have the Gaussian velocity distribution \cite{comment}. 

In Fig.~6 we illustrate how  the granular temperature $T_g$ scales with the energy input rate  and the inelasticity of the particles. We find 
\begin{equation}
\label{temperature}
T_g(\delta V, T, R) = c_0 \zeta^{\lambda}
\end{equation}
where $c_0$ is a constant and $\zeta = \frac{(\delta V)^2/T}{1-R^2}$. Here $(\delta V)^2/T$  is the energy input rate. The exponent $\lambda$ is found to be $0.65 \pm 0.01$. We obtain this scaling form from simulations of $75$ systems with $N=1024$ and  $\rho = 0.16$ by exploring the three dimensional parameter space $(\delta V, T, R)$. Here $\delta V$ sets the values of $0.001$, $0.002$, $0.003$, $0.005$, $0.008$, $T$ the values of $0.0001$, $0.0005$, $0.001$, and $R$ the values of $0.10$, $0.45$, $0.63$, $0.77$, $0.89$. 

\subsection{Correlations in real space}

We have measured the spatial correlation functions by calculating the following quantities:
\begin{equation}
\label{CCrho}
C_{\rho\rho}({\bf r}) = \frac{1}{N}\int d{\bf R} <(\rho({\bf  R}, t) - \bar{\rho})(\rho({\bf  R+r}, t) - \bar{\rho})>
\end{equation}
\begin{equation}
\label{CCvv}
C_{v_{\alpha}v_{\beta}}({\bf r}) = \frac{1}{N}\int d{\bf R} <v_{\alpha}({\bf  R}, t)v_{\beta}({\bf  R+r}, t)>
\end{equation}
\begin{equation}
\label{CCee}
C_{ee}({\bf r}) = \frac{1}{N}\int d{\bf R} <(e({\bf  R}, t) - \bar{e})(e({\bf  R+r}, t) - \bar{e})>
\end{equation}
where $\rho({\bf  r},t) = \sum_{i=1}^{N}\delta({\bf r}-{\bf r}_i(t))$ is the particle density at position ${\bf  r}$ at time $t$, $v_{\alpha}({\bf  r}) = \sum_{i=1}^{N}v_{i \alpha}(t)\delta({\bf r}-{\bf r}_i(t))$ is the momentum density in the $\alpha$--direction and $e({\bf  r},t) = \sum_{i=1}^{N} \frac{1}{2}m{\bf v}_i^2 \delta({\bf r} -{\bf r}_i(t))$  is the energy density (here $v_{i \alpha}$ is the $\alpha$--component of the velocity of the $i$'th particle at time $t$, $\alpha$ and $\beta$ take the values of  $x$ or $y$ and ${\bf r}_i$ is the  position of the $i$'th particle). The $<\cdots>$ means time average and the bar means space average.

From the second rank  correlation tensor for the velocity $C_{v_{\alpha}v_{\beta}}({\bf r})$  we define its transverse ($\perp$) and logitudinal ($\parallel$) components
\begin{equation}
\label{CCpara}
C_{\parallel}(r) = \hat{{\bf r}}_{\alpha} \hat{{\bf r}}_{\beta} C_{v_{\alpha}v_{\beta}}({\bf r})
\end{equation}
\begin{equation}
\label{CCperp}
C_{\perp}(r) = \hat{{\bf r}}_{\perp \alpha} \hat{{\bf r}}_{\perp \beta} C_{v_{\alpha}v_{\beta}}({\bf r})
\end{equation}
where $\hat{{\bf r}}$ and $\hat{{\bf r}}_{\perp}$ are unit vectors parallel and perpendicular to the relative displacement ${\bf r}$ respectively.

For numerical calculations, we first coarse--grain the system into cells of size of about $5\sigma \times 5\sigma$. Fig.~7 shows the plots of correlation functions of $C_{\rho\rho}(r)$, $C_{\parallel}(r), C_{\perp}(r)$ and $C_{ee}(r)$ vs. the distance of $r$. Since the system is isotropic, the correlation functions depend only on the absolute value $r \equiv {\bf r}$.   From Fig.~7(a), we see clearly the long--range correlation in the density fluctuation. For distance less than about $20\sigma$, the correlation is positive while for larger distance the correlation is negative. This is in contrast with the situation of elastic system of $R=1$ where we find $C_{\rho\rho}(r)$ is zero for all $r > 0$. From Fig.~7(b) we can see that both parallel and perpendicular components of velocity are correlated in the long range. For distance above $15\sigma$, the perpendicular velocity correlation is negative.  The results are  different from those for the freely evolving granular medium \cite{Ernst} where there is a characteristic length of vortices determined by the negative minimum of the perpendicular component of velocity correlation. Here there is no negative minimum in  $C_{\parallel}(r)$ for all $r$ less than half of the system size ($L \simeq 70\sigma$). The lack of a negative minimum indicates no characteristic length scale in the system. For the parallel component of velocity correlation, one sees that long--range correlation is also built up for the whole system. In comparison, we find that for elastic system both parallel and perpendicular components are zero for any $r > 0$. These results make us  speculate that our granular system is in a critical state which lacks characteristic length in correlations. 

Fig.~7(c) indicates that there are no correlation (too weak to detect, if any)  in the energy fluctuation, which allows us to name the internal energy as granular temperature in  analogy to the thermodynamics of usual fluids.    

\subsection{Correlations in reciprocal space}

We have also calculated the correlation in the Fourier space. The structure factors are defined as
\begin{equation}
\begin{array}{rcl}
S_{\rho}({\bf k}) & = & \frac{1}{N} \int \int < \rho({\bf r})
            \rho({\bf r}') e^{-{\bf k}\cdot ({\bf r}
            - {\bf r}')} > d{\bf r} d{\bf r}'\\
S_{{\bf v}}({\bf k}) & = & \frac{1}{N} \int \int < {\bf V}({\bf r}) \cdot 
            {\bf V}({\bf r}') e^{-{\bf k}\cdot ({\bf r}
            - {\bf r}')} > d{\bf r} d{\bf r}'
\end{array}
\end{equation}
where $\rho({\bf r}) = \sum_{i=1}^{N} \delta({\bf r} -{\bf r}_i)$ and  ${\bf V}({\bf r}) = \sum_{i=1}^{N} {\bf v}_i \delta({\bf r} -{\bf r}_i)$ are the mass density and momentum density, and $<\cdots>$ means time average. Note that in the above integrals taking {\bf r} = {\bf r}' will  contribute  a constant term (irrepective of ${\bf k}$), which will be dropped in our calculations in the following, since we are not interested in the self--site correlations.  
Thus one expects that both $S_{\rho}({\bf k})$ and $S_{{\bf v}}({\bf k})$ for uncorrelated systems (e.g. ideal gases)  are zero for non--zero ${\bf k}$. From the simulations with $R = 1$ (for which the energy--input procedure is not applied), we obtain that $S_{{\bf v}}({\bf k})$ is indeed zero for non--zero ${\bf k}$. However, due to the finite size of the particles, small but detectable  $S_{\rho}$ was obtained for non--zero ${\bf k}$, which is shown in Fig.~8.  Let $\bar{\rho} w_{12}(r) dV_2$ denote the probability of finding a particle in $dV_2$ given that a particle is in $dV_1$, $S_{\rho}$ can be rewritten as \cite{Landau}:
\begin{equation}
S_{\rho}({\bf k}) = \frac{1}{N} \int d{\bf x}_2 \int d{\bf x}_1  \bar{\rho}^2(w_{12}-1) e^{i{\bf k}\cdot({\bf x}_2 - {\bf x}_1)} 
\end{equation}
Taking the following  hard--core exclusion into consideration,
\begin{equation}
 w_{12} = \left\{ \begin{array} {lcl} \displaystyle 0&, &r \le \sigma\\
                          \displaystyle 1  &,&r > \sigma\end{array}\right.
\end{equation}
where $ r = \left |{\bf x}_2 - {\bf x}_1 \right |$, it is straightforward to obtain the form of $S_{\rho}({\bf k})$
\begin{equation}
\begin{array}{rcl}
\label{square}
S_{\rho}({\bf k}) \simeq -s_0 + s_1 k^2   \hspace{1cm} (\mbox{for} \, \, k << \frac{1}{\sigma})
\end{array}
\end{equation}
where $s_0$ and $s_1$ are $\sigma$--dependent constants.  Fig.~8 indeed shows that $S_{\rho}({\bf k})$ for small values of $k$ obeys Eq.~(\ref{square}). In the following we present results after subtraction of this finite (particle) size effect for the structure factor of mass density. Therefore, by definition, $S_{\rho}({\bf k})$ is zero for non--zero  ${\bf k}$ for elastic systems ($R = 1$). 

Since the system is isotropic,  we obtain the dependence of the structure factors on the magnitude of the wave vector $k$. Let us label the two quantities calculated as described above as $s_{\rho}(k)$ and $s_{{\bf v}}(k)$. Fig.~9(a) (b) show their dependence on $k$ on log--log plot. The linearity indicates that the correlations are of power--law form, indicating no characteristic spatial scales in mass density and velocity density fluctuations. The power--law exponent in  $s_{\rho}(k) \sim k^{-\beta_1}$ is found to keep at a constant value irrespective of $R$ and to our best estimation $\beta_1 = 1.42 \pm 0.06$ while the exponent in $s_{{\bf v}}(k) \sim k^{-\beta_2}$ is dependent of the inelastity.   Fig.~9(c) are the dependence of $\beta_1$ and $\beta_2$ on the restitution coefficient. One should note that the power--law scaling region becomes shorter as $R$ increases. \\ 

\subsection{Distribution of dissipated energy}

Let us now look at  scaling in the energy scale. We focus on all energy dissipation during collisions. The energy change after a collision is $-\delta E$. Fig.~10(a) shows one time series of dissipated energy in  the simulation. Fig.~10(b) shows the distributions of $\delta E$  counted during a long time run for three different restitution coefficients $R=0.6$, $0.3$ and $0.1$. We see  power--law scaling in the energy scale. The upper cutoff of the scaling ($\Delta$) is determined by the requirement that the energy dissipation rate $\gamma \int_0^{\Delta} \delta E P(\delta E) d(\delta E) $ is  equal to the energy input rate ($\delta V^2/T$) where $\gamma$ is the collision rate. We note that the exponent for $P(\delta E) \sim \delta E ^{-\mu} $ decreases as $R$ increases and the upper cutoff ($\Delta$) decreases with increasing $R$. For $R=0.1$, $\mu = 0.91 \pm 0.02$. As  $R$ approaches $1$,  $\mu$ decreases to zero.  

\subsection{Anomalous diffusion}

We have traced the trajectories for some randomly marked particles. Fig.~11(a) is a typical one.  We record the particle positions $\boldmath r$$(t) = (x(t), y(t))$  for a time period of 512 collisions per particle and calculate the power spectra of $x(t)$ and of $y(t)$ (their average is denoted by $S(f)$). Fig.~11(b) and Fig.~11(c) show  $S(f)$ for two different values of $R$. In Fig.~11(b) for $R=0.1$, we see that there are  two regimes separated by $f \sim 100$ which corresponds to a time period of about $60 T$ where $T$ is the heating period. In the high frequency regime, $S(f)$ is proportional to $f^{-2}$, corresponding to the normal diffusion. Since the heating procedure is applied randomly as in Eq.~(\ref{heating}), the diffusive behavior is easily understood in this short time scale regime. Addtionally, we see from  Fig.~11(b) that different behavior exists in the lower frequency regime where $S(f)$ is found to be proportional to  $f^{-(2+\beta)}$ with $\beta$ close to $2$.  This gives rise to the mean square displacement
\begin{equation}
\label{diffusion}
<({\bf r}(t) - {\bf r}(0))^2> \sim t^{1+\beta}.
\end{equation}
The scaling region of this anomalous diffusion becomes shorter as the restitution coefficient increases. This can be seen from Fig.~11(c) which corresponds to $R=0.8$. The power--law scaling in the low frequency regime for $S(f)$ is closely related to the power--law scalings presented in the above subsections. Since the density fluctuation is correlated in space in a self--similar manner as indicated by the power--law scaling in subsection F, the jump of particle, depending on the local density fluctuation, may thus satisfy a self--similar distribution. This has a  link with the L\'evy--flight random walk \cite{Levy}. 

The anomalous diffusion of Eq.~(\ref{diffusion}), obtained in two--dimensional simulations, is apparently similar to the observation by Lewis Fry Richardson in 1926 in the fully developed turbulent flow in three dimensions \cite{Levy,Mandelbrot}. Dimensional analysis gives us  the correlation of the Fourier transformation of velocity  
\begin{equation}
\label{V-V}
<v_k v_{-k}> \sim k^{- \delta}
\end{equation}
with $\delta = (3\beta-1)/(\beta+1)$ which is independent of the dimensionality. If $\beta = 2$, we have the Kolmogorov exponent of $\delta=5/3$. However, the velocity correlation displayed in Fig.~9 for $R=0.1$ shows that $\delta \simeq 1.4$. At present, we do not know whether this discrepancy is intrisic or not. It is possible that an anomalous exponent like the intermittency exponent in three--dimensional turbulent flow \cite{Frisch}  comes into play.

\section{Discussion}
 
The  scaling results obtained in the preceding section are robust to the energy--input procedure we used. We have verified this by using different parameters  for energy--input  ($\delta V$ and $T$) and by using non--periodic energy--input mechanism. We find that the above results hold also irrespective of the average density provided that it is in the  low density regime. For relatively high density, steady state cannot be reached as the particles jam each other in such a way that the system is no longer in a ``gas'' state (i.e., the diffusion constant is approaching zero).

The scaling properties shown in Section III remind us of the concept of self--organized criticality (SOC) \cite{Bak,Sornett}.  In our model we do not need to fine--tune a control parameter to obtain the scalings in space, time and energy scales. The self--organization appears automatically as far as $R < 1$. However, there are several important differences. 
In the common sense of SOC, avalanches  propagate through a medium. The medium  is driven into a critical state  by two opposite processes: external driving  and avalanches' disturbance (due to threshold instability). In our  model, there is no  avalanche. Energy is dissipated instaneously by collision. Three processes are responsible for the evolution of the system: external driving, energy dissipation and self--evolving due to the velocity field. The last factor is lacking in the SOC of the sandpile model.

Our model is also different from turbulence of ordinary fluids \cite{Taguchi}. In a fully developed turbulence, there is a cascade phenomenon which is the origin of the scaling behaviors such as the Kolmogorov scaling. The energy is input to the system in a macroscopic scale, and then transfered into  shorter scales and finally dissipated at a microscopic length scale due to viscosity. In our model each particle gains kinetic energy randomly. Therefore, there is no macroscopic length (except the system size) asscociated with the energy input. In this sense the Reynolds number is negligiblly small in our system. 
We have obtained an anomalous diffusion of a test particle, which is similar to the Richardson law. However, it should be noted that our system is in two dimensions and that scaling properties in two dimensions are quite different from those in three dimensions in ordinary fluid turbulence. 
For further theoretical studies for the scalings obtained here, one might need to take into account a new non-dimensional parameter \cite{us2} controlled by dissipation due to inelastic collision. 

In conclusion we have shown by a well--designed model that essential differences exist between granular systems and  elastic ones. By self--organization our system reaches a critical state where there are no characteristic spatial and temporal scales in correlations, and no characteristic energy scale in dissipation.

\begin{center}
{\bf ACKNOWLEDGMENTS}
\end{center}

This work was supported by
Grant--in--Aid of Ministry of Education, Science and Culture of Japan. 
G. P. thanks the Japan 
Society for the Promotion of Science.

\begin{figure}
\caption{Energy relaxation starting from a random initial configuration for  a system with $N=1024$,  $\rho = 0.16$ and  $R = 0.8$. Vertical axis is the kinetic energy per particle ($\frac{1}{2}m <v^2>$ with $m=2$). Best fit using Eq. (7) is also shown with $E(\infty) = 1.21 \times 10^{-3}$, $A = 4.42 \times 10^{-4}$ and $b' = 0.38$. 
}
\label{Fig. 1}

\end{figure}
\begin{figure}
\caption{(a) Real time vs. total number of collisions  for system with $N=1024$ and $\rho = 0.16$. The restitution coefficients are $R=0.5$, $0.4$, $0.3$, $0.2$ respectively from top to bottom. (b) Collision rate ($\gamma$) vs. degree of dissipation $\epsilon = 1 - R^2$. The curve is the best fit using Eq. (8) to the numerical data with $\gamma_0 = 505.045$, $\gamma_1 = 130.064$ and $c=0.635$.
}
\label{Fig. 2}
\end{figure}

\begin{figure}
\caption{
(a) Typical configuration of particles. This graph was obtained from simulations for system with  $N=1024$, $\rho = 0.16$ and $R=0.2$. Arrows represent velocities. (b) Visualization of places where dissipation has taken place during a time interval of 20 collisions per particle. When a pair of particles collide, we mark a data point ($\diamond$) at the middle point between the centers of the pair. Here $N=1024$, $R=0.1$, and $\rho = 0.16$. 
}
\label{Fig. 3}
\end{figure}

\begin{figure}
\caption{
(a) Power spectrum $P(f)$ of the  time--dependent coordinates of dissipation places where collision occurs.  The straight line is the best fit to the numerical data, giving $\alpha = 0.49 \pm 0.01$.  (b) Dependence of the exponent $\alpha$ on $R$.
}
\label{Fig. 4}
\end{figure}

\begin{figure}
\caption{
Velocity distribution for system with  $N=4096$, $\rho = 0.16$ and $R=0.6$. Vertical axis is the number of counts for velocity square in a bin between $v^2$ and $v^2 + \delta v^2$ with the bin size $\delta v^2 = 1.77 \times 10^{-5}$.
}
\label{Fig. 5}
\end{figure}

\begin{figure}
\caption{
Granular temperature $T_g$ vs. $\zeta = \frac{(\delta V)^2/T}{1-R^2}$.  The straight line is the best fit to the numerical data, $T_g = c_0 \zeta^{\lambda}$ with $c_0 = 4.97 \times 10^{-3}$ and $\lambda=0.65$.
}
\label{Fig. 6}
\end{figure}

\begin{figure}
\caption{
(a) Density--density correlation function $C_{\rho\rho}$; (b) Parallel ($\diamond$) and  perpendicular ($+$) components of velocity--velocity correlation functions; (c) Energy--energy correlation function. Results are obtained for system with $N=1024$, $\rho = 0.16$ and $R=0.4$.  Horizontal axis is in unit of particle diameter $\sigma$. 
}
\label{Fig. 7}
\end{figure}

\begin{figure}
\caption{
Structure factors $S_{\rho}(k)$ for elastic system ($R=1$) of $N=4096$ and $\rho = 0.16$. Best fit using the form of Eq. (18) is also shown with $s_0 = 0.55$ and $s_1=1.35 \times 10^{-6}$.
}
\label{Fig. 8}
\end{figure}

\begin{figure}
\caption{
(a) Structure factor $S_{\rho}(k)$; (b) Structure factor $S_{{\bf v}}(k)$. Power--law decaying exponents are $\beta_1 = 1.39 \pm 0.02$ and $\beta_2 = 1.40 \pm 0.04$ for system with $R=0.1$, $N=4096$ and $\rho = 0.16$.   Straight lines are best fits to the numerical data in the range of k from $2\pi$ to 100. The length scale corresponding to a wave vector $k$ is $2\pi/k$. k=100 corresponds to a length of about $8\sigma$. (c) Dependence of exponents $\beta_1$ ($\diamond$) and $\beta_2$ ($+$) on $R$.
}
\label{Fig. 9}
\end{figure}

\begin{figure}
\caption{
(a) Energy change ($-\delta E$) at collision for a process with $N=1024$, $\rho = 0.16$ and $R=0.1$. (b) Distribution of dissipated energy $P(\delta E)$ for three systems with $N=1024$, $\rho = 0.16$ and different $R$ ($R=0.6, 0.3, 0.1$ from right to left). }
\label{Fig. 10}
\end{figure}

\begin{figure}
\caption{
(a) A  trajectory for one marked particle taken from simulation  with $N=1024$, $\rho = 0.16$ and $R=0.1$. (b) Power spectrum $S(f)$ of the time--dependent coordinates of a marked particle vs. frequency $f$ for system with $N=1024$, $\rho = 0.16$ and $R=0.1$. The two lines have slopes of $-4.17$ and $-1.97$ respectively. (c) Same as (b) but for $R=0.8$. The two lines have slopes of $-4.16$ and $-1.93$ respectively. }
\label{Fig. 11}
\end{figure}


\begin{references}
\bibitem{comment} We have found that when our system is exposed to an uniaxial field (gravity) the velocity distribution satisfies no longer Gaussian distribution for the whole system. However, Gaussian distribution is still valid for any  part of the system with height $\delta h$ as log as $mg \delta h << \frac{1}{2} m <{\bf v}^2>$. This is very different from the situation of elastic systems where the Boltzmann distribution governs the density changes but the average kinetic energy is $k_BT$ anywhere in the system.
\end{references}
\end{document}